# Framework for Version Control & Dependency Link of Components & Products in Software Product Line

Faheem Ahmed, Luiz Fernando Capretz, Miriam Capretz

Department of Electrical & Computer Engineering
University of Western Ontario
London Ontario, Canada, N5A 5B9
{sgraha5, lcapretz, mcapretz}@uwo.ca

*Abstract: -* Software product line deals with the assembly of products from existing core assets commonly known as components and continuous growth in the core assets as we proceed with production. This idea has emerged as vital in terms of software development from component-based architecture. Since in software product line one has to deal with number of products and components simultaneous therefore there is a need to develop a strategy, which will help to store components and products information in such a way that they can be traced easily for further development. This storage strategy should reflect a relationship between products and components so that product history with reference to components can be traced and vise versa. In this paper we have presented a tree structure based storage strategy for components and products in software product line. This strategy will enable us to store the vital information about components and products with a relationship of their composition and utilization. We implemented this concept and simulated the software product line environment.

## 1. Introduction

A software product line is a set of software-intensive systems sharing a common, managed set of features that satisfy the specific needs of a particular market segment or mission and are developed from a common set of core assets in a prescribed way [1]. Software product line is a collection of systems sharing a managed set of features constructed from a common set of core assets [2]. The objective of a software product line is to reduce the overall engineering effort required to produce a collection of similar systems by capitalizing on the commonality among the systems and by formally managing the variation among the systems [3]. Research [4,5,6,7] has been done on the process definition and associated relevant activities of software product line. Although software product line is gaining popularity over the passage of time due to economical impact [8], there has not been a great deal of research in establishing appropriate approach for managing version control and link dependency of components and products in software product line.

## 1.1 Problem Definition

A product consists of multiple components and composition of the products describes how and which components are used to develop it. Similarly a component can be used in multiple products within a software product line. There is a need to define a strategy to represent the software product line activity in terms of relationship among the components and the products. Every component should be traceable with reference to its utilization and version history. Similarly every product should completely describe its

1782

composition with reference to components. This strategy should work at the software product line level not only at the individual component and product level. This will help everybody in the development and management levels to get access this information. In this paper we have presented a tree structure based storage strategy for components and products in software product line. This strategy will enable us to store the vital information about components and products with a relationship of their composition and utilization.

## 1.2  Tree Structure Scheme

A tree t is a finite non-empty set of elements. One of these elements is called the root, and the remaining elements (if any) are partitioned into trees, which are called the sub trees of t [9]. The general tree can have a start node called root node and it can have any number of children. Each child serves as parent for the next level of the tree.

Figure 1.1 represents the root node of the tree and we named it "Software Product Line". This tree can be used to represent all the activities performed during software product line development with reference to components and products. The root node "Software Product Line" has a left child as "Core Assets Repository" and right child as "Product". The "Core Asset Repository" will store the information about various components developed during the software product line development activity. The right child "Product" will store information about all the products developed during the software product line development activity.

Figure 1.2 represents the development tree of software product line. The left child of the root "Software Product Line" represents the core assets repository having children C1, C2, C3 and further C4 and C5. The children C4 and C5 of C1 and C2 represent the later versions of C1 and C2 respectively. The right children of the root represent the products developed with its own left child core asset repository, which represents the composition of product P1 and P2 individually. The figure illustrates that the composition of P1 is C1, C2 and C3, while the composition of P2 is C4 and C5.

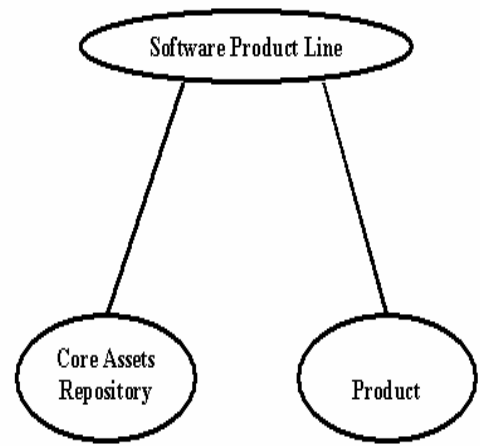

**Figure 1.1 General Structure of Software Product Line Tree**

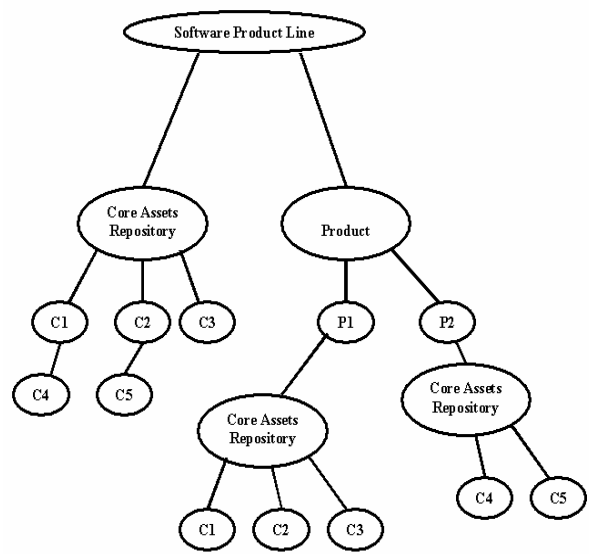

**Figure 1.2 Development Tree of Product Line**

```
public void TraverseTree(Object present_node)
{
 DefaultMutableTreeNode nodetosearch=(DefaultMutableTreeNode)present_node;
 int ii=SPL_Tree.treePanel.treeModel.getChildCount(nodetosearch);
    for(int i=0;i<ii;i++)
       {
TraverseTree((Object)SPL_Tree.treePanel.treeModel.getChild(nodetosearch,i));
       }
}
```

**Listing 1.1 Code Segment for Traversing the Software Product Line Tree**

## 1.3 Interpretation of the Development Tree

The traversing of the tree gives us important information about the software product line activity in conjunction with core components and products developed. Traversing the tree can retrieve the following information:

- Information about the core asset repository of the software product line can be obtained by traversing the left child of the root node of software product line tree. This information will enable developers to know how many and what components are present to be utilized. Figure 1.2 describes that C1, C2, C3, C4 and C5 are present in the core asset repository and can be used in any further development of a product.
- Information about the versions of a component present in the tree can be observed, for example Figure 1.2 describes that C1 and C2 has later versions as C4 and C5 respectively.
- Information about the products developed can be obtained by traversing the right child of the root node. The information will allow us to know how many products have been developed. Figure 1.2 describes that two products P1 and P2 are developed yet.
- Traversing at each Product node can give us information about the structural composition of the product by visiting the left child of each product. For example in Figure 1.2 the product P1 is composed of C1, C2 and C3 while product P2 is composed of C4 and C5 respectively.
- If we further expand the development tree keeping the same topology, the tree can give us information about various versions of the products as well.

## 2. Implementation of the Approach

The implementation of the approach is carried out in Java to simulate the tree structure of the development tree. The JTree class of java swing components is used to represent the tree. Various methods of the JTree are implemented to perform the basic tree operations like insert, delete, and update nodes. Figure 1.3 represents the basic architecture of the implementation approach. The application is divided into different layers, which are tree operations, graphical user interface, XML handling routines, database manipulation routines and search and retrieval routines. The application reads and stores information about development tree into structured XML file which is being handled by XML handling routines layer. The associated information about each component and product like developed date, author, release date, test date, testing result and author information etc is stored into the database and is being handled by database manipulation routines layer. The GUI layer facilitates display and user interaction with the application. The tree operation layer provides basic tree manipulation operations such as insert delete and update node. To traverse the entire tree a simplified recursive approached is used, Listing 1.1 is the code segment. This method takes a node as an argument and finds the number of children for the present node. This method recursively call itself for each child of the node. The method recursively call itself until the node has no child. Whenever a new product is added a product node along with one left child as "Core Asset Repository" according to the scheme is added to the tree. Listing 1.2 shows the XML code for the development tree. XML (extensible Markup Language) is a standard language for defining structured of text-based data, and XML technologies are the framework for data transactions such as data management, data exchange, data conversion, data collection and data distribution [10]. It constructs the root node as "Software Product Line" and left child node as "Core Asset Repository" and right child node as "Product". The same scheme works for all inserted product nodes later in the XML structure for example Product "P1" and Product "P2". Listing 1.3 shows the DTD structure of XML file for the development tree of software product line. A DTD specifies what elements may occur and how the elements may nest in an XML [11]. Listing 1.3 shows that root element is "Software Product Line" and it contains two children "Core Asset Repository" and "Product". The Core Asset Repository can further have zero or many Core Assets. And the Product will have two children Core Asset

Repository and further version of product. Figure 1.4 is the snap shot of the GUI screen of the implementation. It uses JTree component of Java to render the XML based tree.

```
<?xml version="1.0" ?>
<Software_Product_Line>
 <Core_Asset_Repository>
        <C1><C4></C4></C1>
        <C2><C5></C5></C2>
        <C3></C3>
 </Core_Asset_Repository>
 <Product>
        <P1>
           <Core_Asset_Repository>
                <C1></C1>
                <C2></C2>
                <C3></C3>
           </Core_Asset_Repository>
        </P1>
        <P2>
           <Core_Asset_Repository>
                <C4></C4>
                <C5></C5>
           </Core_Asset_Repository>
        </P2>
 </Product>
</Software_Product_Line>
```

**Listing 1.2 XML Code of Software Product Line Development Tree**

```
<!DOCTYPE SoftwareProductLine [
<!ELEMENT
 SoftwareProductLine (CoreAssetRepository , Product+)>
<!ELEMENT          CoreAssetRepository
(CoreAsset*)>
<!ELEMENT CoreAsset (#PCDATA)*>
<!ELEMENT Product (CoreAssetRepository ,Products*)>
<!ELEMENT          CoreAssetRepository
(CoreAsset*)>
<!ELEMENT CoreAsset (#PCDATA)*>
<!ELEMENT Products (#PCDATA)*>
]>
```
**Listing 1.3 DTD Structure of Software Product Line Development Tree**

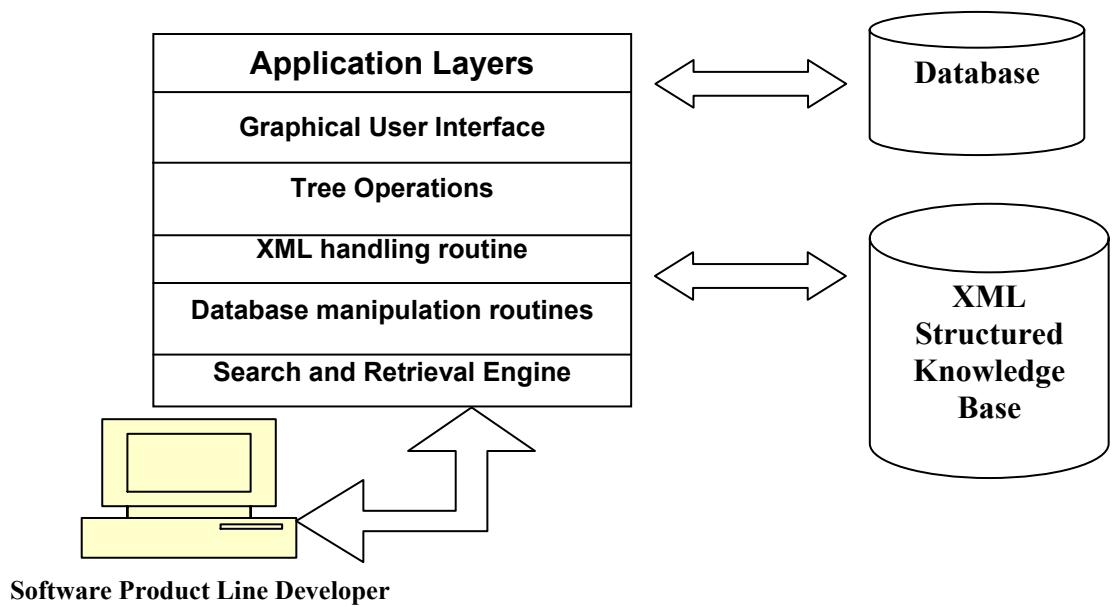

**Figure 1.3 Implementation Architecture of the Approach**

## 3.  Future Research and Conclusion

Software Product Line requires a scheme, which can handle the information about core assets and product simultaneously to enhance the productivity of the developers. The implementation of our proposed approach shows that it handles the complex structure of information about components and products simultaneously with their defined association. It gives information about components and products at all levels of software product line activity ranging from an individual product or component to the software product line itself. We are working on developing a CASE Tool for software product line. This framework for version control and link dependency of components and products will be used as a layer in the software to store and retrieve information about components and products under development or developed. The developers will be able to get information about various components already developed and used in other products. The management will be able to extract information about the status of the product line.

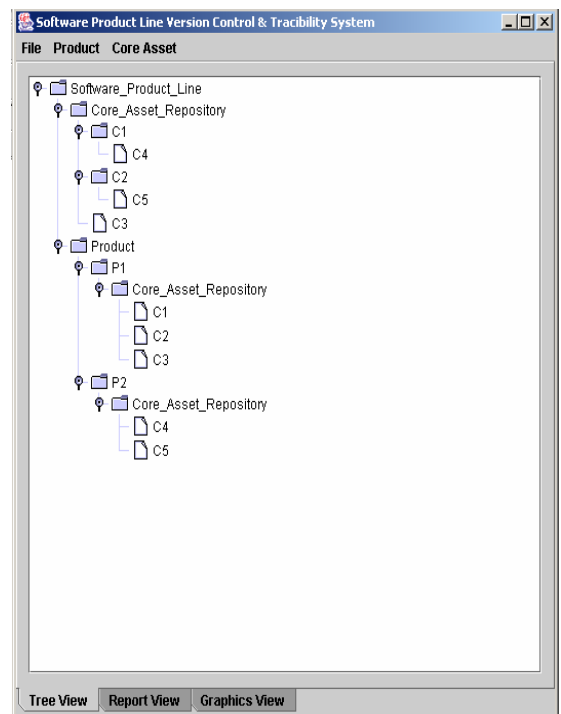

**Figure 1.4 Snap Shot of the GUI Screen of the implementation**

## 4.0  References

[1] P. Clement, L. Northrop, "Software Product Lines, Practices and Pattern", Addison Wesley, 2002.

[2] L. Bass, P. Clements, R. Kazman , "Software Architecture in Practice". Addison Wesley, 1998.

[3] Krueger, C.W, "Software product line reuse in practice" Application-Specific Systems and Software Engineering Technology, 2000. Proceedings. Of The 3rd IEEE Symposium , 24-25 March 2000 Pages:117 – 118

[4] Bass, L.; Clements, P. and Kazman, R. (1998) Software Architecture in Practice, Addison-Wesley

[5] John, I. and Schmid, K. (2001) Product Line Development as a Rationale Strategic Decision, Proceedings of the International Workshop on Product Line Engineering, Seattle, pp.31-35.

[6] Knauber, P. and Succi, G. (2000) Software Product Lines: Economics, Architectures and Applications, Proceedings of the International Conference on Software Engineering, Limerick, pp. 814 – 815.

[7] Linden, V.D.F. (2002) Software Product Families in Europe: ESAPS & Café Projects, IEEE Software, pp. 41-49.

[8] Buckle, G.; Clements, P.; McGregor, J.D.; Muthig, D. and Schmid, K. (2004) Calculating ROI for Software Product Lines, IEEE Software Magazine, Volume 21, No 3, pp. 23-31

[9] S. Sahni, "Data Structure, Algorithms, and Applications in Java" McGraw Hill, 2000.

[10] Ishitani,Y , "Document transformation system from papers to XML data based on pivot XML document method" Document Analysis and Recognition, 2003. Proceedings. Of The Seventh International Conference on, 3-6 Aug. 2003 Pages: 250 - 255 vol.1

[11] Kotsakis, E, Bohm. K, "XML Schema Directory: a data structure for XML data processing" Web Information Systems Engineering, 2000. Proceedings of The First International Conference on, Pages: 62 - 69 vol.1, 19-21 June 2000